# Influence of applied voltage and electrical conductivity on underwater pin-to-pin pulsed discharge

C. Rond [1,a], J.M. Desse[2] , N. Fagnon[1], X. Aubert[1], A. Vega[1] and X. Duten[1]

[1]LSPM - CNRS UPR3407, Université Paris 13, Villetaneuse, 93430, France

[2]ONERA - DAAA and LMFL, CNRS FRE 2017, 5 Boulevard Paul Painlevé, BP 21261, 59014 LILLE Cedex, France

**Abstract.** A parametric study of an underwater pulsed plasma discharge in pin-to-pin electrode configuration has been performed. The influence of two parameters has been reported, the water conductivity (from 50 to 500 µS/cm) and the applied voltage (from 6 to 16 kV). Two complementary diagnostics, time resolved refractive index-based techniques and electrical measurements have been performed in order to study the discharge propagation and breakdown phenomena in water according to the two parameters. A single high voltage of duration between 100 µS and 1 ms is applied between two 100 µm diameter platinum tips separated by 2 mm and immersed in the aqueous solution. This work, which provides valuable complementary results of paper [1], is of great interest to better understand the mechanisms of initiation and propagation of pin-to-pin discharge in water. For low conductivity (from 50 to 100 µS/cm) results have confirmed two regimes of discharge (cathode and anode) and the increase of the applied voltage first makes the breakdown more achievable and then favors the apparition of the anode regime. For 500 µS/cm results have highlighted cathode regime for low applied voltage but a mixed regime (anode and cathode) for high applied voltage.

## 1. Introduction

Previous observations indicate that pre-breakdown and breakdown phenomena in water are extremely varied and complex. In particular it has been shown that the formation of discharges is very dependent on the conductivity of the water and the magnitude of the applied voltage, these parameters change the energy input and so the kinetics of the process [2, 3]. Many studies have reported the influence of the applied voltage for discharge in dielectric liquids [4-7] but only few works concerns water discharge [8]. Different discharge modes have been observed depending on the amplitude of the applied voltage pulse: slow streamers showing a hemispheric or a bush-like shape are related to low amplitude voltage whereas fast streamers appear above a certain threshold voltage (according to the geometry of the set up) and can be considerably longer [2, 8-11]. Moreover the effect of the applied voltage has been investigated on discharge parameters such as the breakdown time [12, 13] or the OH radical density [14].

It is also expected that the electrical conductivity of the solution is a crucial parameter of the set up since the solvated ions modify the space charge electric field in the liquid. As a consequence it would determine the initiation mechanisms of the

a Author to whom correspondence should be addressed. Electronic mail: rond@lspm.cnrs.fr

electrical discharge, depending on either the capacitor or the resistor liquid behavior [15]. Indeed, if the discharge ignition is related to the formation of a bubble by ohmic heating (and not by electron injection from the electrodes), the conductivity of the liquid plays a significant role [16]. Few works have reported a large influence of the solution conductivity on the direct liquid phase electric discharge for different geometries mainly as pin-to-plane and plane-to-plane [2, 15, 17-21]. The increase of the solution conductivity results in higher discharge current (higher concentration of ions), a shorter pulse duration (faster current dissipation between the gaps) and a shortening of the streamer length (faster compensation of the space charge electric field on the head of the streamer). This involves higher power density in the channel resulting in higher electron density, larger UV radiation power and a strong effect on chemical species formation. But research to date has not highlighted a clear relationship between the solution conductivity and the discharge physics [15, 22]. On the one hand it has been shown that the increase of the conductivity has no influence on discharge parameters such as the discharge inception voltage [20] or the time lag to breakdown [13]. On the other hand, other studies found that the conductivity has a significant influence on the discharge mechanisms, but the results show controversy: higher conductivity leads to either increase [18] or decrease the threshold voltage of the breakdown [15].

Most of these works have used the pin-to-plane electrodes system for underwater pulsed discharge experiment but few studies concern symmetric pin-to-pin configuration. Recently, the authors have reported the analysis of pre-breakdown and breakdown phenomena for underwater pin-to-pin discharge [1]. Keeping all the experimental parameters constant, two regimes of discharge have been identified regarding its structure, its velocity and its propagation way. Cathode regime has been defined as bush-like structures propagating slowly from the cathode to the anode whereas anode regime has depicted a filamentary structure emerging faster from the anode. It is to notice that the cathode regime can be subdivided into 2 cases as Case (1) which corresponds to partial discharge and Case (2) which involves complete discharge (the channels span the electrode gap). Main results have shown that the cathode regime is characterized by the simultaneous apparition of a gas phase at the cathode and a transient current overlapping a RLC current. On the contrary, the anode regime does not involve a transient current and no vaporization at the cathode has been observed.

This work has provided new insights into plasma-liquid physics and has also raised some issues to be discussed. The complete understanding of discharge mechanisms in water requires experimental observations varying experimental conditions such as the electric field or the liquid conductivity. With a similar approach than in [1], the authors report synchronized measurements on a single discharge using both time-resolved optical techniques (schlieren imaging and shadowgraphy) and electrical measurements in order to allow valuable investigations of the effect of the conductivity and the applied voltage on the discharge initiation and propagation.

## 2. Experimental set-up

A short description of the experimental set up is presented in Fig.1 since the lector would find a detailed depiction in [1]. The reactor (100x45x50) mm is made of PMMA and holds two fused silica windows and two electrodes facing each other. Platinum 100 μm-diameter electrodes are placed in a horizontal pin-to-pin configuration with a gap distance of 2 mm. The pulse generator consists of a 1 nF capacitor constantly charged by a 30 kV high voltage power supply and discharged using a fast high voltage solid-state switch. Positive high voltages mono-pulses with rise time of about 30 ns, adjustable duration from 100 μs to 1 ms and amplitude from 6 to 16 kV are produced. The electrical measurements are achieved using a high voltage probe connected to the high voltage electrode and a coaxial current shunt (R = 10 Ω) connected to the low potential electrode. Both signals are recorded simultaneously with a 1 GHz Digital Storage Oscilloscope (DSO-Lecroy HDO9104). The source light of the schlieren optical set up is a 300 W Xenon light source and the signal is monitored by a high speed camera (Phantom V1210). Videos have been recorded using two different parameter sets: an exposure time of 0.91 μs and a widescreen resolution of 128x32 pixels which allows 571 500 frames per second; an exposure time of 0.56 μs with a widescreen resolution of 128x128 pixels which allows 240 600 frames per second. It is worth noting that the camera allows monitoring both variation of the refractive index and strong light emissions. The experiments were performed in a mixture of de-ionized water and sodium chloride in order to vary the conductivity (σ=50, 100 and 500μS/cm).

Special attention was paid to maintain constant the initial conditions of the discharge, in particular the time between two measurements was high enough to avoid residual bubbles. It is worth noting that a minimum of dozen of measurements have been carried out for each experimental condition in order to ensure the reproducibility of the observed phenomena and to provide reliable analysis of the results.

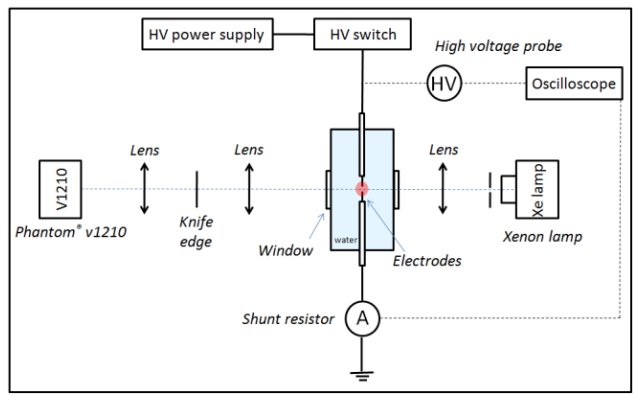

Fig. 1 Schematic diagram of the experimental setup (top view)

## 3. Influence of solution conductivity

Experiments have been carried out for three different conductivities (50, 100 & 500 µS/cm). The corresponding Maxwell relaxation time (ε/σ) ranges from 14 ns (σ=500µS/cm) to 140 ns (σ=50µS/cm), which characterizes the resistive behavior of the liquid during the discharge [23]. As well as the conflicting results found in the literature [15, 18, 20], this work reports inconclusive experimental results about the influence of the conductivity on the discharge inception voltage since it is equal to 7, 9 and 6 kV for 50, 100 and 500 µS/cm respectively. This first result highlights the complexity of plasma-liquid and the strong influence of the conductivity on the ignition process.

### 3.1. Discharge characterization

Time-resolved refractive index-based techniques and electrical measurements have been used to characterize the discharge in the liquid. According to criteria defined in [1] and mentioned previously, experiments can be divided into two different regimes and 3 cases. A statistical analysis over a hundred of experiments shows the distribution of the cases and the number of breakdowns according to the conductivity. As an example for 12 kV (Tab.1), we report no Case (1) and a distribution of Case (2)/Case (3) as 83/17 for 50µS/cm, 90/10 for 100µS/cm and 68/32 for 500µS/cm. The two lower conductivities show similar ratio between cathode and anode regimes but the higher conductivity makes the anode regime more probable.

| REGIME | CATHODE | | | | | ANODE |
|---|---|---|---|---|---|---|
| | Case 1 | Case 2 | | | | Case 3 |
| (%) | | 1 Bk | 2 Bk | 3 Bk | 4 Bk | |
| 50 µS/cm | 0 | 83 | | | | 17 |
| | | 0 | 18 | 51 | 31 | |
| 100 µS/cm | 0 | 90 | | | | 10 |
| | | 0 | 32 | 45 | 23 | |
| 500 µS/cm | 0 | 68 | | | | 32 |
| | | 71 | 29 | 0 | 0 | |

Tab.1 Probability of cases and number of breakdowns in relation with the conductivity (U=12 kV)

For each case the interesting features of schlieren measurements are compared according to the 3 conductivities. These phenomena are related to the analysis of voltage and current waveforms. As discussed in [1] the electrical traces are similar to waveforms of RLC circuit with an overdamped response, the corresponding RLC current can be overlapped by a transient component. The results are presented from Fig.2 to Fig.4 according to the cases.

**Cathode regime – Case (1) – Fig.2:** The discharge features are quite similar for 50 and 100 μS/cm whereas significant differences appear for 500 μS/cm (Fig.2(a)). For low conductivities, low variations of the refractive index are first observed at both electrodes indicating an increase of the liquid temperature. Then bush-like dark channels form at the cathode resulting from the vaporization of the water, contrarily to the anode. The solution conductivity influences the channels length since the maximum length is about twice higher for 50 μS/cm than for 100 μS/cm. Due to the higher conductivity, the high ion concentration involves a fast current dissipation in the solution and therefore shorter channels [15, 18]. At 500 μS/cm a gas phase appears fast at both electrodes, at the cathode the shape of the gas phase is also bush-like whereas at the anode the gas phase expansion is radial.

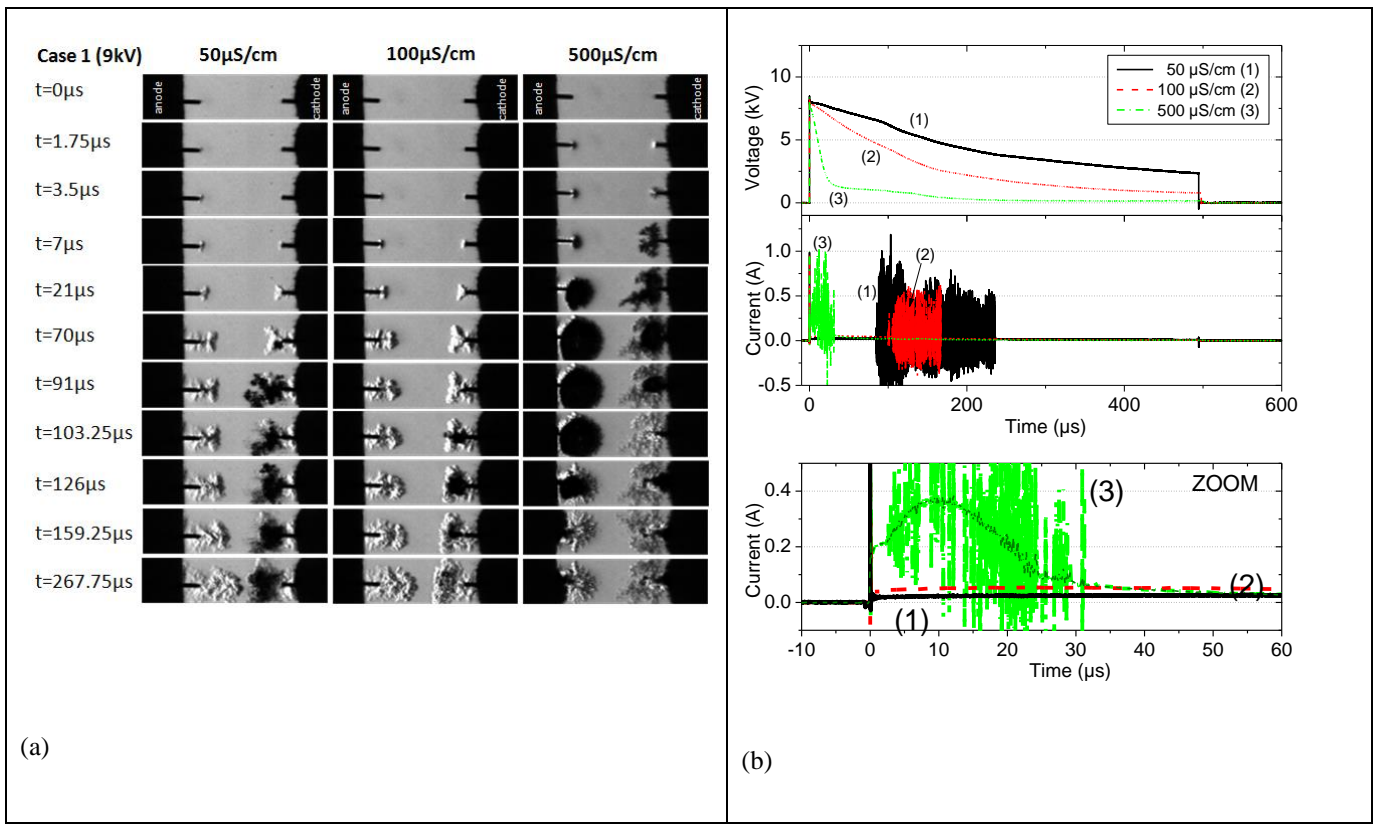

*Fig 2: Cathode regime – Case (1) (9 kV) for 3 conductivities (50, 100 & 500 μS/cm) (a) Time-resolved schlieren images (b) electrical waveforms (and a zoom of the current)*

The voltage pulse width shortens with the increase of the conductivity because of more efficient charge transfer (Fig.2(b)). This variation confirms that our system acts as a RLC circuit with an overdamped response. For a series resonant circuit, the quality factor Q depends inversely on the resistance, and so proportionally to the conductivity. It is known that as the quality factor Q decreases the slower decay mode dominates the system's response resulting in a slower system. Despite the transient current overlap, the RLC current can be estimated by smoothing the current waveform. As expected the maximum value of the RLC current increases with the conductivity being equal to 20 mA, 50 mA and 350 mA for 50, 100 and 500 µS/cm respectively. The RLC current monitored for 500 µS/cm highlights a two-stages increase, (a first plateau is reached at 200 mA before the maximum value of 350 mA - see zoom in Fig.2(b)), which is not clearly observed for low conductivities. Moreover the transient current is temporally well-defined for 50 and 100 µS/cm but for 500 µS/cm the transient current is spread over the RLC current making it difficult to define the beginning of the transient signal (delay time) and its duration.

**Cathode regime – Case (2) - Fig.3:** For all tested conductivities, the discharges behave similarly (Fig.3(a)). Initial variations of the fluid density are observed at each electrode tip. Then a gas phase expands quite radially and slowly from the anode whereas gas channels propagate with bush-like structure from the cathode to the anode. The propagation of the gas phase depends on the solution conductivity, in particular the expansion of the gas phase at the anode is more important with higher conductivity. Then channels cross the interelectrode gap which leads to the breakdown phenomena. As a consequence an intense light is emitted from the gas phase. Next the gas phase expands in a radial shape and several secondary emissions can be observed in the gas phase (*e.g.* at 141.75µs-50µS/cm and at 71.75µs-100µS/cm). The conductivity has an influence on the number of breakdowns, they are highly present for low conductivity solutions but very scarce at high conductivity. As an example for 12 kV (Tab.1), a statistical analysis shows that for low conductivities 2, 3 and 4 breakdowns are observed whereas at 500 µS/cm only 1 and 2 breakdowns have been monitored.

As for Case (1) the voltage pulse width shortens with the increase of the conductivity (Fig.3(b)). For low conductivities (50&100 µS/cm), the current is also composed by transient and RLC currents, the RLC currents reach 100 and 300 mA respectively. However for 500 µS/cm, the transient current no longer exists and the RLC current shows a maximum at about 700 mA.

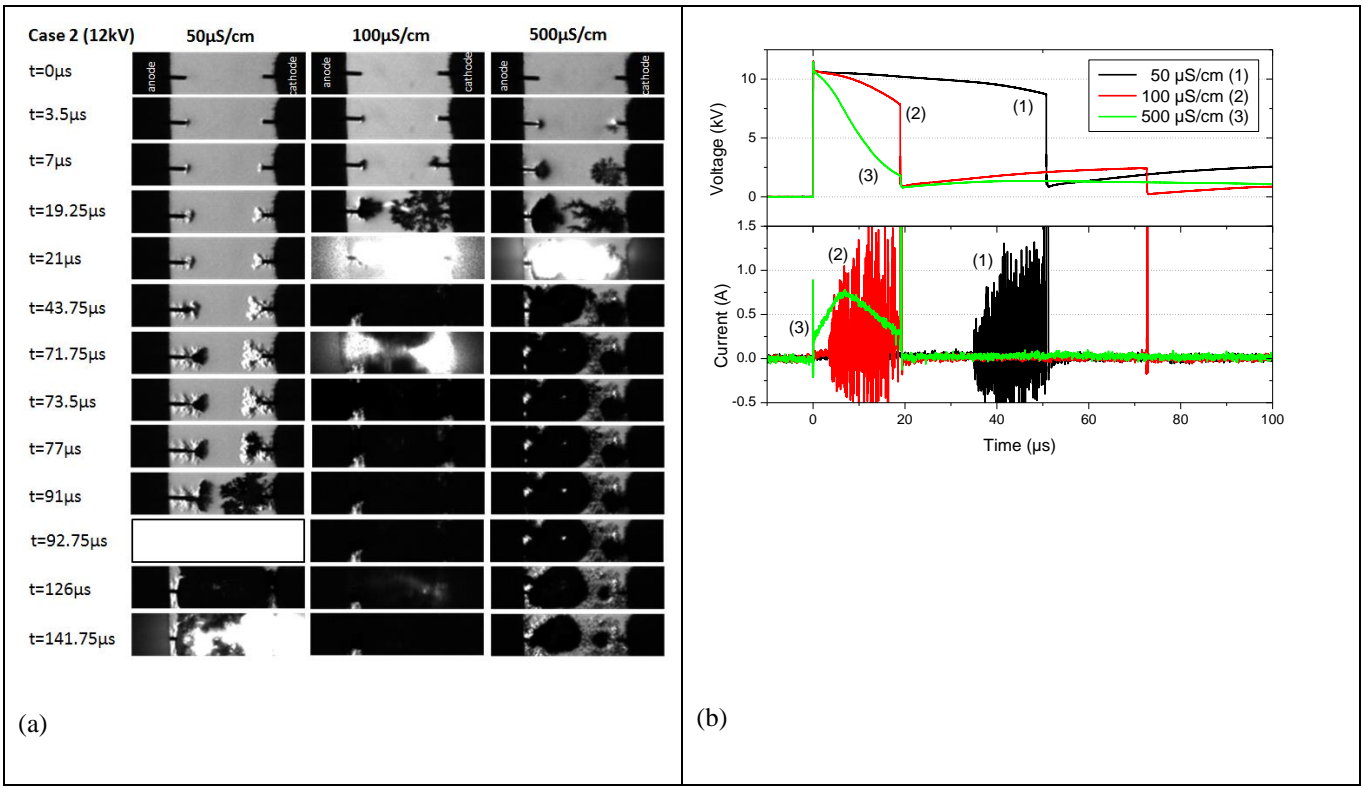

*Fig 3: Cathode regime - Case 2 (12 kV) for 3 conductivities (50, 100 & 500 μS/cm) (a) Time-resolved schlieren images - the white image means that the camera is saturated by the plasma emission (b) Electrical waveforms*

**Anode regime – Case (3) - Fig.4:** For all tested conductivities a filamentary structure initiates and propagates from the anode toward the cathode at a very high velocity ($v_3$>400 m/s) (Fig.4(a)). The time resolution of the camera limits the study of the pre-breakdown phenomena but it appears that for 500 μS/cm a gas phase is also formed at the cathode before the breakdown. Following the first breakdown, several secondary breakdowns are systematically observed. As previously, when the conductivity increases, the voltage pulse width shortens and the RLC current increases (Fig.4(b)). The current waveforms are strongly dependent on the conductivity. For 50 μS/cm, the RLC current is very low and a weak transient current may appear. For 100 μS/cm, RLC current is about 100 mA whereas no significant transient current is monitored. For 500 μS/cm, the current looks like a "noisy" RLC current, we can assume that it is composed by RLC current overlapped by a weak transient current. These results highlight that the conductivity has a strong influence in anode regime characteristics.

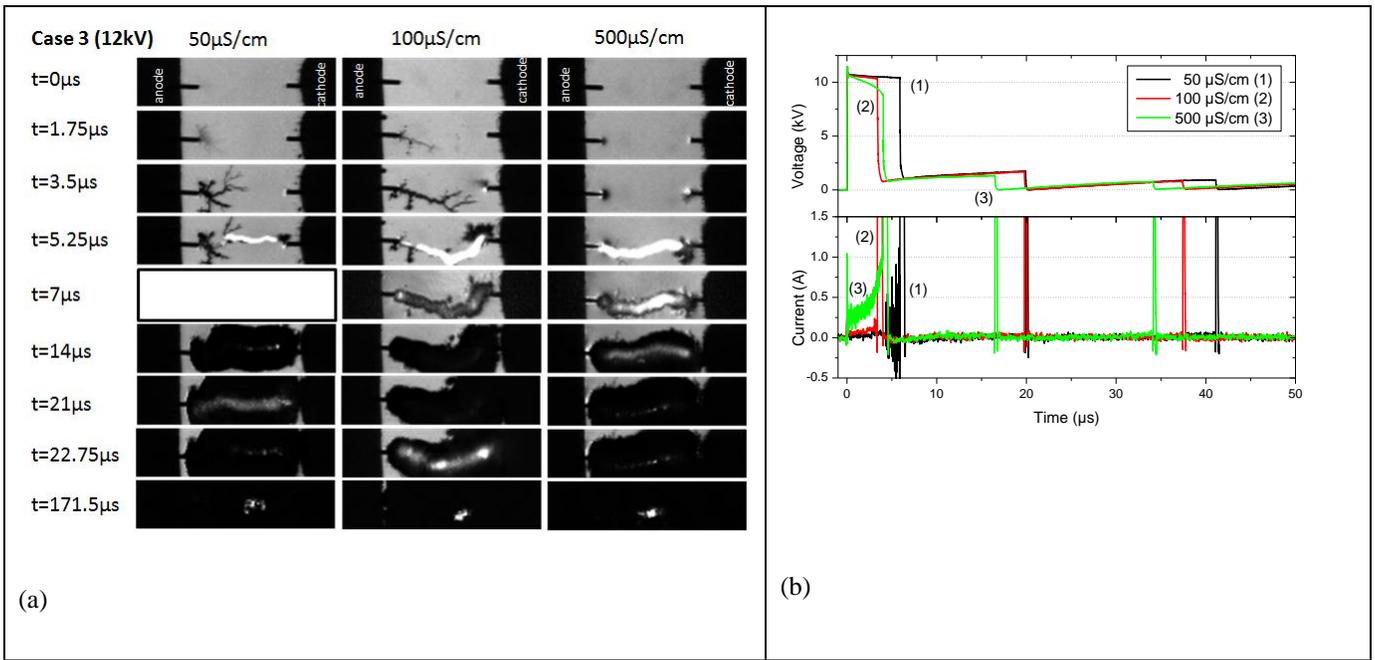

*Fig 4: Anode regime – Case (3) (12 kV) for 3 conductivities (50, 100 & 500 µS/cm) (a) Time-resolved schlieren images - the white image means that the camera is saturated by the plasma emission (b) Electrical waveforms*

As a summary, whatever the regime of the discharge, the influence of the conductivity on the voltage is similar - the voltage pulse width shortens and the RLC current increases with the conductivity rise, as expected. But the transient current show discrepancies between low conductivities (50-100 µS/cm) and high conductivity (500 µS/cm) for the two regimes. This result is consistent with the statistical analysis (Tab.1) and the refractive index images which have highlighted differences between results obtained at low and high conductivities.

### 3.2. Pre-breakdown analysis

As defined in [1], the delay time corresponds to the time needed for the appearance of the gas phase at the cathode and those of the transient current, whereas the duration corresponds to the duration of the transient current and those of the gas phase expansion. Results obtained in the cathode regime (each measurement is based on a dozen of experiments) are reported on Fig.5 for the three conductivities. As explained above, current waveforms at 500 µS/cm do not show any well-defined transient current so the corresponding electrical measurements are not reported on Fig.5.

The influence of the conductivity on the delay time is not significant since the dispersion of the measurements is important. However we observe that this dispersion decreases when the conductivity increases. The influence of the conductivity on the duration is far more significant for Case (1) than for Case (2) since the duration of Case (1) decreases when the conductivity increases whereas no significant influence of the conductivity is observed for Case (2).

Higher conductivity of the solution leads to faster current dissipation as already mentioned, and also higher current flow (as shown in Fig.2) which facilitates the vaporization process by ohmic heating. If Joule effect plays a significant role in discharge initiation, the increase of the conductivity facilitates the discharge formation, as observed for Case (1). However the non-dependence of Case (2) duration in conductivity confirms that the gas phase production does not only depend on Joule effect as already mentioned in [1].

For low conductivities, a good agreement is obtained between the imaging and electrical measurements which confirms that transient current and vaporization at the cathode are related whereas this correlation cannot be verified for 500 µS/cm since transient current is not well defined.

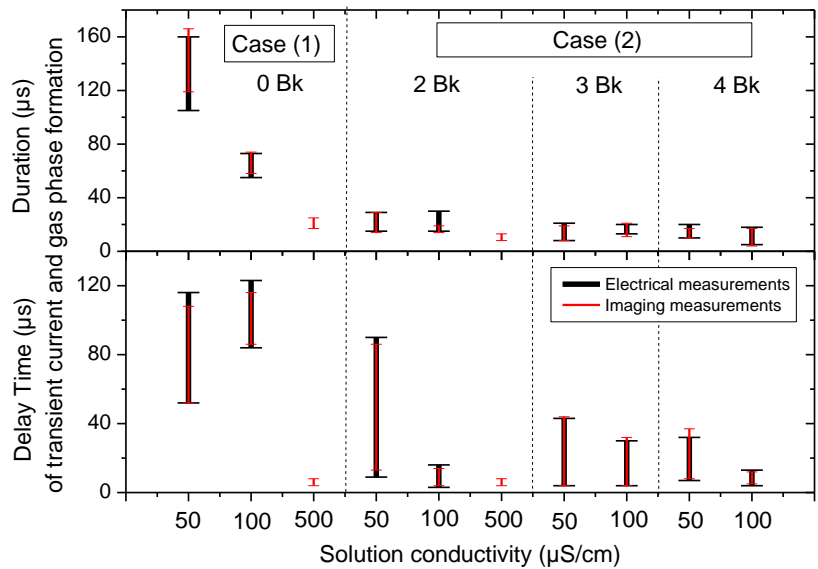

**Fig 5: Influence of the solution conductivity on the delay time and duration of the pre-breakdown for Cathode regime:**
**Case (1)** -U=9kV - *(0 breakdown)* and **Case (2)** - U=12kV- *(2, 3 and 4 breakdowns).*

### 3.3. Analysis of the breakdown phenomena

Fig.6 represents the breakdown voltage in relation with the current peak value for the three conductivities at U=12 kV. Firstly no effect of the conductivity is observed for secondary breakdowns of Case (2) and Case (3) (unfilled symbols), the electric measurements corresponding to these breakdowns are lined up along a straight line passing through the origin. The experiments satisfy the equation $U^{bk}=R.I^{pk}$ with R ≈150 Ω for all conductivities. These results suggest that mechanisms responsible for secondary breakdowns are similar whatever the conductivity, so the parameter R does not depend on the electrical properties of the fluid but on those of the gas phase. It means that the secondary breakdowns occur in the gas phase that has been previously formed by the 1st breakdown.

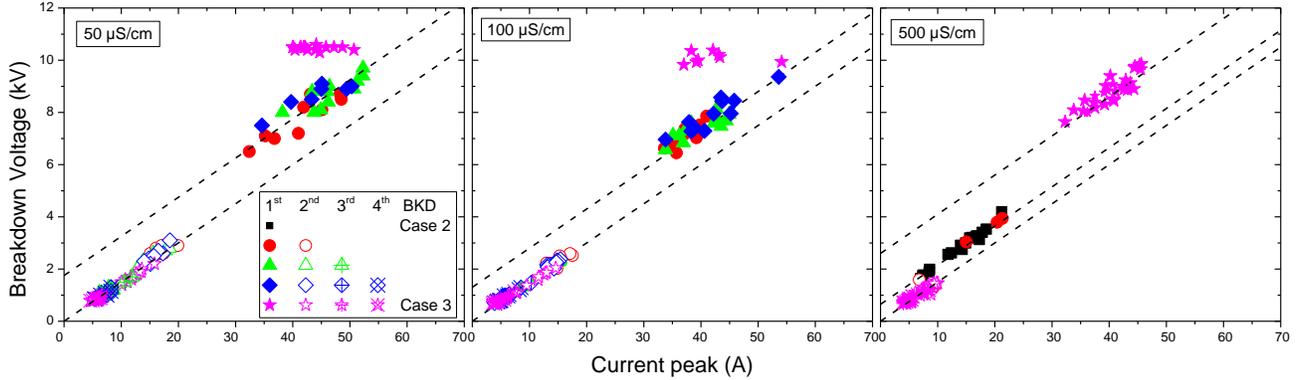

**Fig 6:** *Breakdown voltage according to current peak for the conductivities (50,100,500 µS/cm) for U=12 kV. Regarding the symbols signification, the geometry refers to the cases and number of breakdowns and the filling refers to the rank of the breakdown*

Secondly the electrical measurements of first breakdowns of Case (2) (filled symbols except star) are lined up along a line not passing through the origin and parallel to the previous one ($U^{bk}$=E+R.$I^{pk}$). As previously the slope of the line is equal to R≈150 Ω and does not depend on the conductivity, so the current peak of the first breakdown is due to charges transfer in a gas phase. This result confirms the formation of a gas phase prior to the first breakdown in the frame of the cathode regime. However the value of the intercept E decreases when the conductivity increases, being equal to $E_{50}$=1.75 kV, $E_{100}$ = 1.3 kV and $E_{500}$ = 0.65 kV, which confirms that this parameter is a characteristic of the liquid properties [1].

Furthermore, for the same applied voltage (U=12 kV) we report that the first breakdown voltage and the first current peak of Case (2) decreases when the conductivity increases (especially visible between 100 and 500 µS/cm). The charge transferred to the liquid during the pre-breakdown and the breakdown has been estimated by the voltage drop at the measurement capacitor C=1 nF (Q=C.ΔU). The values of charges reported in Fig.7 show that the distribution of the initial charge (10.8 µC) between the pre-breakdown and the 1st breakdown depends on the conductivity. The charge delivered during the pre-breakdown increases with the conductivity. This variation is mainly due to the increase of the RLC current (Fig.3), since the pre-breakdown duration is not changing significantly (Fig.5). As a summary, the higher concentration of ions makes the solution a better electrical conductor so more charges can be initially transferred in the liquid. As a consequence, when more charge is transferred during the pre-breakdown, a lower amount of charge is available for the 1st breakdown which induces lower breakdown voltage and peak current (Fig.6) and leads to less additional breakdowns (as shown in Tab.1).

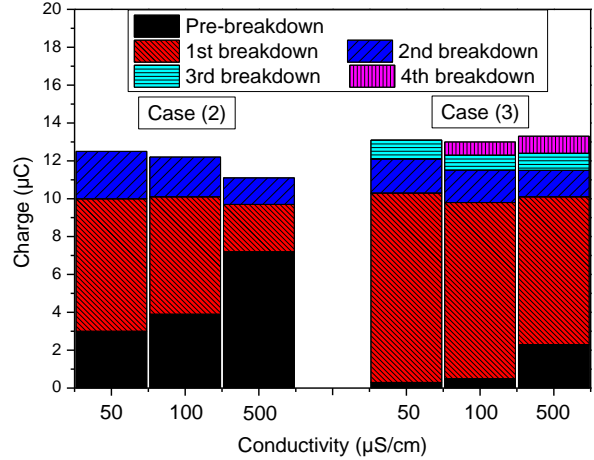

**Fig.7** Mean values of charge (in µC) transferred during the pre-breakdown (Pre-Bk) and the breakdowns according to the conductivity for an applied voltage U= 12 kV for Case (2) (only results of 2 breakdowns are reported) and Case (3)

Finally the first breakdowns of Case (3) (star symbols) show very different features between low and high conductivities (Fig.6). For low conductivities (50&100 µS/cm), the measurements are lined up along a horizontal line according to the equation $U^{bk}=E'$, with E'= 10.5 kV for U=12 kV (it is to notice that E' does not depend significantly on the conductivity for a given applied voltage). For high conductivity (500 µS/cm) the experiments are satisfying the equation $U^{bk}=E'+R.I^{pk}$ with R≈150 Ω which is characteristic of the propagation of the discharge in a gas phase. This is consistent with schlieren images and electrical measurements presented on Fig.4 which have reported a gas phase propagating from the cathode and a low transient current. These results have been previously defined as characteristics of Case (2). However the intercept of the equation (E'=2.6 kV) is higher than those measured for 1st breakdown of Case (2) (E=0.65kV) and schlieren measurements have reported filaments propagating from the anode to the cathode which has defined Case (3) discharge. As a result we can conclude that the measurements carried out for 500 µs/cm and high applied voltage present features of both Case (2) and Case (3). The discharge mechanisms for high conductivity and high applied voltage are driven by both cathode and anode processes.

## 4. Influence of the applied voltage

The applied voltage is varied from 6 to 16 kV while the other parameters are kept constant, the conductivity of the water solution is set to 100 µS/cm. The measurements have shown that the threshold of the applied voltage required for initiating a discharge, which corresponds to either the gas phase formation at the cathode or a transient current monitoring, is 9kV. From this value, the increase of the applied voltage, *i.e.* the injected energy, changes the distribution of the regimes defined previously. As summarized in Tab.2, for 9 kV only the cathode regime has been observed, divided into 5% of Case (1) (no gap bridge) and 95% of Case (2) (gap bridge). For higher values of applied voltages, Case (1) is no more observed. From 10 to 11 kV, only the Case (2) of the cathode regime has been observed. Then by increasing the applied voltage, Case (3) appears and its proportion increases since we report a ratio Case (2)/Case (3) equals to 90/10 for 12 kV, 72/28 for 14 kV and 42/58 for 16 kV. The increase of the applied voltage first makes the breakdown more achievable and then favors the apparition of the anode regime.

| | Case 1 | Case 2 | | | | | | Case 3 |
|---|---|---|---|---|---|---|---|---|
| | | *1 Bk* | *2 Bk* | *3 Bk* | *4 Bk* | *5 Bk* | *≥6bk* | |
| **9kV** | 5 | 95 | | | | | | 0 |
| | | *70* | *30* | *0* | *0* | *0* | | |
| **10kV** | 0 | 100 | | | | | | 0 |
| | | *42* | *50* | *8* | *0* | *0* | | |
| **12kV** | 0 | 90 | | | | | | 10 |
| | | *0* | *32* | *45* | *23* | | | |
| **14kV** | 0 | 72 | | | | | | 28 |
| | | *0* | *0* | *42* | *48* | *8* | *2* | |
| **16kV** | 0 | 42 | | | | | | 58 |
| | | *0* | *0* | *6* | *52* | *33* | *9* | |

***Tab.2:*** *Probability of cases and breakdowns according to the applied voltage (in %)*

As discussed previously, Case (2) analysis is of particular interest since the pre-breakdown is characterized by simultaneous transient current and gas phase formation at the cathode. It is worth noting that the electrical traces do not show significant modifications with the increase of the applied voltage since the maximum voltage obviously increases, as well as the current peak, but the intensities of RLC and transient currents do not change notably (Fig. 8).

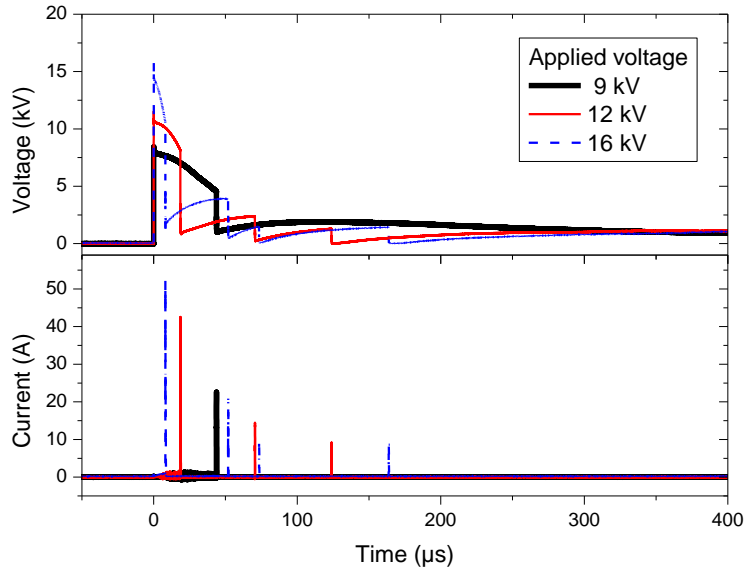

Fig. 8 Voltage and current signals of a pin-to-pin discharge in water (σ=100μS/cm) monitored for applied voltages of 9kV, 12kV and

16kV (Cathode regime - Case (2))

Fig 9 reports the delay time and the duration of both the transient current and the gas phase formation measured by the electrical and the imaging diagnostics respectively (dozen of experiments reported for each condition). The applied voltage does not have a significant influence on either the delay time or the duration. It means that the mechanisms responsible for ignition and propagation of Case (2) do not strongly depend on the electric field.

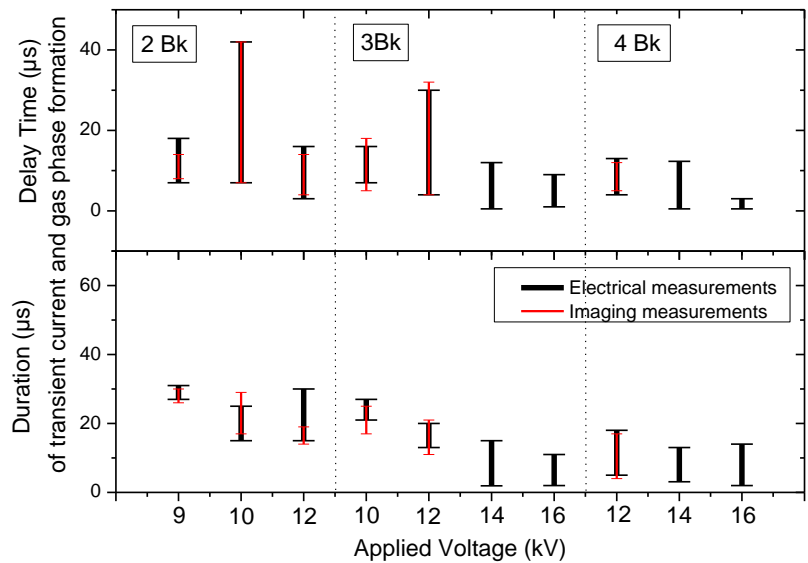

Furthermore for Case (2) the experiments have been distinguished according to the number of breakdowns during the discharge (from 1 to 6). A statistical analysis (based on one hundred of experiments) shows that the increase of the applied voltage favors high number of breakdowns (Tab.2). These results can be related to the pre-breakdown characteristics, the RLC current and the pre-breakdown duration (which is defined as the sum of the delay time and the duration of the transient current reported on Fig.9), and they can be interpreted in terms of injected charge. Calculations highlight that the distribution of the initial charge into pre-breakdown and breakdown depends on the applied voltage. The increase of the applied voltage favors the charge delivered during the 1$^{st}$ breakdown whereas the pre-breakdown charge remains almost constant. As an example for 3 breakdowns, charge delivered during the 1$^{st}$ breakdown represents 45%, 50%, 56% and 60% of the initial charge for 10, 12, 14 and 16 kV respectively (Fig.10). A high electric field leads to a higher amount of charge involved in the breakdown which could induce a higher excitation/ionization of the species and then favors more additional breakdowns.

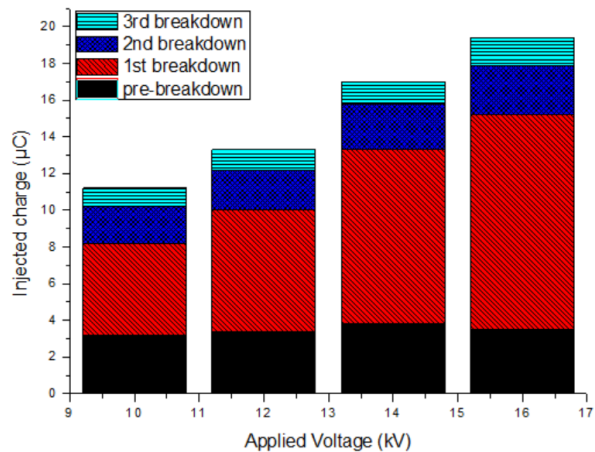

**Fig.10** *Mean values of charge (in μC) transferred during the pre-breakdown and the breakdowns according to the applied voltage, for Case (2) (3 breakdowns)*

It has to be noted that the increase of the charge delivered during the breakdown corresponds to the increase of the peak current. The peak current is related to the breakdown voltage by the breakdown (I/V) characteristic as presented in Fig.11. It is to notice that measurements of secondary breakdowns are not represented in Fig.11 since no significant influence has been observed, only first breakdowns of Case (2) and Case (3) are represented.

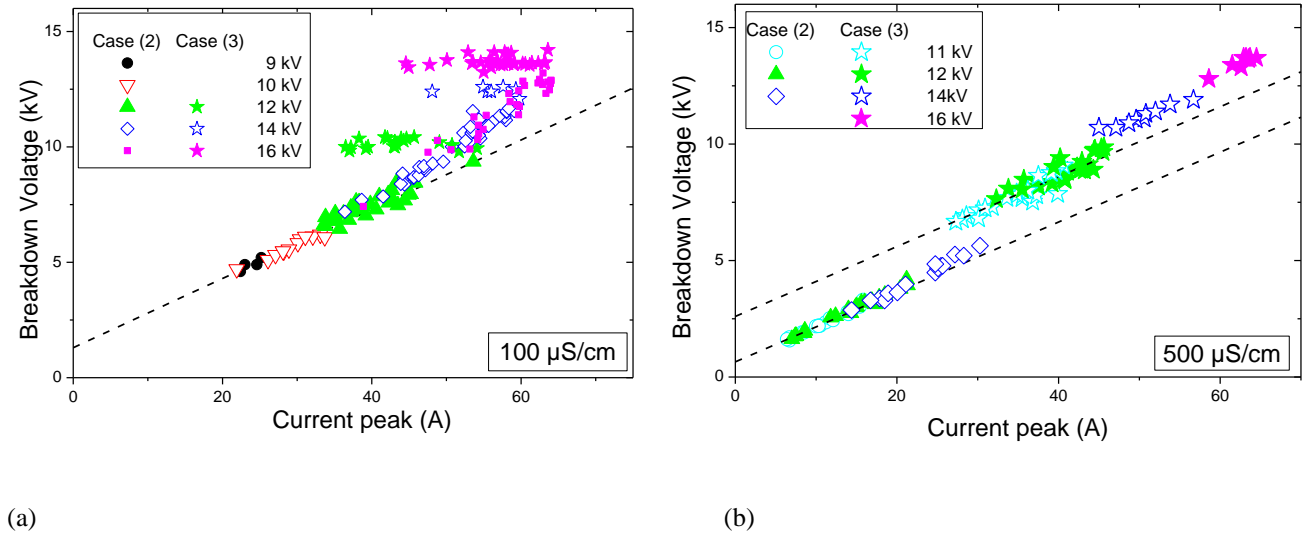

(a)                                                                                      (b)

**Fig. 11:** *Breakdown voltage according to current peak for different applied voltages with (a) σ=100μS/cm and (b) σ=500μS/cm. Only first breakdown of Case (2) and Case (3) are represented (independently of the number of breakdowns)*

The mean values of the breakdown voltage and the peak current for 1$^{st}$ breakdown increase with the applied voltage. For 100 μS/cm (Fig.11(a)), results of cathode regime are lined up along one unique straight line ($U^{bk}=E+R.I^{pk}$) for applied voltage up to 12 kV. For higher applied voltages, measurements reported on the breakdown (I/V) characteristic depart slightly from this trend. If considering that the breakdown electrical properties can be modeled by the simple Ohm law, the measurements also verify the equation $U^{bk}=E+R.I^{pk}$ but with different values of E and R. For low voltages, we have estimated E=1.3 kV and R=150 Ω as previously reported whereas for high voltage, E=0 V and R=200 Ω. However there is a possibility that the breakdown electrical properties are modeled by a more complex equation as a power law. Even more results are necessary to go further on this subject, these new results highlight that channels observed by schlieren technique is a complex medium, certainly not only made of neutral gas but containing excited species that properties are strongly dependent on the electric field. Moreover the variation of the threshold voltage, which is related to the vaporization necessary for the first breakdown, confirms that the formation of the gas phase involve not the only Joule effect but also mechanisms dependent on electric field (as Fowler-Nordheim electron emission).

For anode regime (Case (3)), first breakdowns are lined up along a horizontal straight line ($U^{bk}=E'$) (Fig.11(a)) and the applied voltage has a strong influence on the breakdown voltage as: E'$_{12kV}$=10.2 kV, E'$_{14kV=}$ 12.5 kV and E'$_{16kV}$ = 14 kV. The breakdown voltage increases with the applied voltage. This result confirms that Case (3) involves a threshold voltage necessary to initiate the breakdown. Since the initiation controls the breakdown, the breakdown voltage strongly depends on the initial value of the applied voltage.

The influence of the voltage for an aqueous solution at 500 µS/cm is also reported since it shows particular results which deserve to be emphasized (Fig.11(b)). Whereas the influence of the applied voltage on Case (2) cannot be verified (no Case (2) has been monitored below 11 kV and above 14 kV), the distribution of Case (3) is modified by the applied voltage. First breakdowns of Case (3) are all lined up along straight lines ($U^{bk}=E'+R.I^{pk}$) of slope equal to R=150 Ω, this value being not modified by the applied voltage, but the threshold voltage depends on the applied voltage, increasing from $E'_{12kV}=$ 2.5 kV to $E'_{16kV}=$4.1 kV. The effect of the voltage on the breakdown (I/V) characteristic confirms that the discharges obtained at high voltage for 500 µS/cm present features of both Case (2) and Case (3), defining a mixed regime.

## 5. Conclusion

Despite a unique set of experimental conditions pin-to-pin discharges in aqueous solution propagates according two different regimes (involving 3 cases) [1]. The cathode regime is characterized by a gas phase initiating at the cathode, propagating slowly in a bush-like structure towards the anode and related to the measurements of a transient current. The anode regime, which does not show any gas phase at the cathode or transient current, is defined by filaments propagating faster from the anode to the cathode. The mechanisms that cause these differences are not well understood. This paper presents experimental results obtained by two complementary diagnostics, electrical and refractive index measurements, by varying the solution conductivity and the applied voltage. The results have been obtained over at least a dozen of experiments which provide reliable evolutions of the considered parameters. Regarding the dispersion of the experiments, a dedicated approach involving a massive statistical study would be of great interest to improve the discharge analysis.

On the one hand, the results have shown that the increase of the applied voltage first makes the breakdown more achievable and then favors the apparition of the anode regime. It could be interpreted in terms of injected charge since the increase of the applied voltage results in the increase of the injected charge, especially during the first breakdown process which also favors more additional breakdowns. The influence of the applied voltage on the breakdown characteristics ($U^{bk}$, $I^{pk}$) highlight that channels observed by schlieren technique is a complex medium, certainly not only made of neutral gas but containing excited species that properties are strongly dependent on the electric field.

On the other hand, the variation of the solution conductivity also provides valuable results. We have confirmed that our system acts as a RLC circuit with an overdamped response since the voltage pulse width shortens and the RLC current increases with the increase of the conductivity. These new results also prove that a gas phase propagates from the cathode to the anode in the frame of the cathode regime, and that the anode regime involves a threshold voltage necessary to initiate the breakdown. We have highlighted strong discrepancies about the discharge characteristics between measurements performed at low conductivity (50 & 100 µS/cm) and high conductivity (500µS/cm). For the low conductivities, the characteristics of the two regimes as described previously have been observed whereas for 500 µS/cm some variations have been noticed:

(i) Discharges obtained at high conductivity and high applied voltage correspond to a mixed case between Case (2) and Case (3), they are driven by both cathode and anode processes.

(ii) Case (1) (cathode regime) presents gas phase formation at the anode and a more complex current signal.

(iii) Case (2) (cathode regime) does not present any transient current but a gas phase at the cathode. This result suggests that the transient current cannot be related to the only vaporization process, since in the case of high conductivity the formation of the gas phase seems not to be accompanied by charges creation (no transient current).

As other discrepancy between Case (1) and Case (2), the results of the present paper show that the Joule effect plays a more important role for discharge initiation of Case (1) than Case (2). For the latter case the vaporization process at the cathode does not depend only on Joule effect. All these results confirm the necessity to consider separately the two cases of the cathode regime.

The discharge in liquids depends on many different parameters; the conductivity and the applied voltage are ones of the most important since their influence is significant, but also different for the initiation and propagation of the discharge. Estimation of injected charge has shown that the increase of the solution conductivity favors the deposition of the charge during the pre-breakdown whereas the increase of the applied voltage changes mainly the charge deposited during the first breakdown.

These results confirm the complexity of plasma-liquid and the necessity to multiply the diagnostics in order to provide a full description of the involved mechanisms.

**ACKNOWLEDGMENTS**

The authors wish to thank Université Paris 13 for the financial support and Vision Research Company for the large assistance regarding the use of their equipment.